\documentclass[10pt,conference]{IEEEtran}

\usepackage{amsmath,amssymb,amsthm}
\usepackage{graphicx}
\usepackage{cite}
\usepackage{tikz}
\usepackage{quantikz}
\usepackage{braket}
\usepackage{algorithm}
\usepackage{algpseudocode}
\usetikzlibrary{positioning}

\begin{document}

\title{High Precision Qubit-Efficient Variational Continuous Optimization via Amplitude Estimation}

\author{
\IEEEauthorblockN{Parth Danve}
\IEEEauthorblockA{
University of Connecticut\\
Storrs, CT, USA\\
parth.danve@uconn.edu $|$ parthdanve39@gmail.com
}
}

\maketitle

\begin{abstract}
Optimization of continuous-variable objectives on standard gate-based quantum computers via variational algorithms such as QAOA is typically approached by first discretizing each decision variable into a finite binary representation. This increases qubit requirements and restricts solution precision through fixed-resolution encodings. We propose a qubit-efficient variational framework for continuous optimization that instead encodes each decision variable into the squared amplitude or equivalently, the measurement probability of a single qubit state. This removes explicit discretization from the variable representation while remaining entirely within the standard qubit circuit model unlike methods like CV-QAOA employing qumode based hardware to achieve the same. To read out encoded variables, we propose using amplitude estimation rather than naive sampling or tomographic reconstruction, with the goal of improving precision scaling for continuous-value recovery. We outline how amplitude-estimation error propagates to decision-variable error and then to objective-value error under standard regularity assumptions, suggesting a distinct width-versus-precision tradeoff relative to discretized approaches. In particular, the framework replaces the logarithmic increase in qubits needed for finer binary precision with a constant cost of one qubit per decision variable, while shifting accuracy requirements into the estimation procedure. We position this approach relative to traditional discretized variational formulations, and argue that it provides a promising new direction for continuous optimization on standard qubit architectures.
\end{abstract}


\section{Introduction}
\label{sec:intro}

Continuous optimization underlies a broad class of problems in engineering, finance, and machine learning, where the goal is to minimize a real-valued objective $f(\mathbf{x})$ over $\mathbf{x} \in \mathbb{R}^n$ subject to bound constraints. Variational quantum algorithms (VQAs) \cite{cerezo2021variational}, and in particular the Quantum Approximate Optimization Algorithm (QAOA) \cite{farhi2014qaoa}, have emerged as a leading paradigm for heuristic optimization on near-term noisy intermediate-scale quantum (NISQ) devices \cite{preskill2018nisq}. These algorithms employ a parametrized quantum circuit as an ansatz, iteratively optimizing circuit parameters via a classical outer loop to minimize an objective encoded in a quantum Hamiltonian.

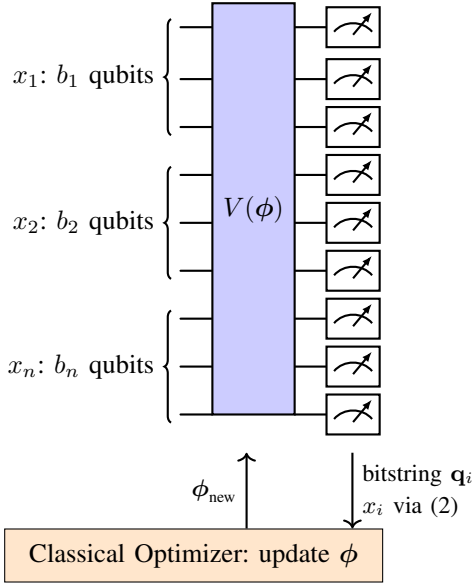
\begin{figure}[t]
\centering
\begin{tikzpicture}
    % Quantum circuit node
    \node[anchor=center] (circuit) {
        \begin{quantikz}[row sep=0.3em, column sep=1.2em]
            \lstick[3]{$x_1$: $b_1$ qubits} & \gate[12, style={fill=blue!20}]{V(\boldsymbol{\phi})} & \meter{} \\
             & & \meter{} \\
             & & \meter{} \\
            \lstick[3]{$x_2$: $b_2$ qubits} & & \meter{} \\
             & & \meter{} \\
             & & \meter{} \\
            \lstick[3]{$x_n$: $b_n$ qubits} & & \meter{} \\
             & & \meter{} \\
             & & \meter{} 
        \end{quantikz}
    };
    % Classical optimizer box below
    \node[draw, rectangle, fill=orange!20, below=1.0cm of circuit, minimum width=5cm, minimum height=0.7cm] 
        (optimizer) {Classical Optimizer: update $\boldsymbol{\phi}$};
    % Right arrow: bitstrings going down (right side)
    \draw[->, thick] 
        ([xshift=2.1cm]circuit.south) -- 
        node[right, align=left, font=\small] {bitstring $\mathbf{q}_i$\\[2pt]$x_i$ via \eqref{eq:decoding}} 
        ([xshift=2.1cm]optimizer.north);
    % Left arrow: new parameters going up (left side)
    \draw[->, thick] 
        ([xshift=0.7cm]optimizer.north) -- 
        node[left, font=\small] {$\boldsymbol{\phi}_{\text{new}}$} 
        ([xshift=0.7cm]circuit.south);
\end{tikzpicture}

\caption{Standard discretized variational circuit for $n$ continuous 
decision variables. Each variable $x_i$ is encoded into a dedicated 
register of $b_i$ qubits as in \eqref{eq:bitwidth}, and a joint 
variational ansatz $V(\boldsymbol{\phi})$ acts across all registers. 
Measurement bitstrings are decoded via \eqref{eq:decoding} and passed 
to a classical optimizer which updates $\boldsymbol{\phi}$ iteratively. 
Total qubit count grows as $\sum_{i=1}^n b_i$, increasing logarithmically 
with the required precision $1/\eta_i$ per variable.}
\label{fig:discrete_circuit}
\end{figure}

When applied to continuous-variable objectives, however, standard gate-based quantum computers present a fundamental representational mismatch: qubits are inherently binary, yet the decision variables are continuous. The dominant approach resolves this by \emph{discretizing} each continuous variable into a finite binary string \cite{farhi2014qaoa, tan2021qubit, glos2022space}. Concretely, for a variable $x_i \in [L_i, U_i]$ with range $R_i = U_i - L_i$ and desired precision $\eta_i$, the number of qubits required is
\begin{equation}
    b_i = \left\lceil \log_2\!\left(\frac{R_i}{\eta_i}\right) \right\rceil,
    \label{eq:bitwidth}
\end{equation}
and the variable is recovered from a bitstring $(q_i^{(0)}, q_i^{(1)}, \ldots, q_i^{(b_i-1)}) \in \{0,1\}^{b_i}$ via the decoding map
\begin{equation}
    x_i = L_i + \frac{R_i}{2^{b_i}} \sum_{k=0}^{b_i - 1} q_i^{(k)}\, 2^k.
    \label{eq:decoding}
\end{equation}
For an $n$-variable problem, the total qubit count grows as $\sum_{i=1}^{n} b_i$, which scales logarithmically with the inverse precision $1/\eta_i$ for each variable. The resulting circuit structure is illustrated in Fig.~\ref{fig:discrete_circuit}, where each decision variable occupies a dedicated register of qubits and a variational ansatz $V(\boldsymbol{\phi})$ acts across all registers jointly.

This discretization imposes two compounding costs. First, the qubit 
requirement grows with the desired precision, making fine-grained 
continuous optimization increasingly resource-intensive on near-term 
hardware. Second, the encoding in \eqref{eq:decoding} introduces an 
irreducible \emph{discretization error} of at most $R_i / 2^{b_i}$ 
per variable, placing a hard floor on solution quality that cannot be 
overcome by deeper circuits or more shots.

An alternative paradigm for avoiding discretization altogether is 
continuous-variable (CV) quantum computing, in which information is 
encoded in the quadrature amplitudes of bosonic modes (qumodes) rather 
than qubits \cite{weedbrook2012gaussian}. CV-QAOA \cite{verdon2019cvqaoa} 
and its experimental realization \cite{enomoto2023cvqaoa} demonstrate 
that continuous optimization can be performed natively on CV hardware 
without any binary encoding. However, this approach requires fundamentally 
different hardware from the standard qubit model, limiting its applicability 
to photonic or other bosonic platforms which remain in relatively early 
stages of development compared to standard qubit architectures.

In this paper, we propose a third path: a qubit-efficient variational 
framework for continuous optimization that avoids binary discretization 
\emph{while remaining entirely within the standard qubit circuit model}. 
The key idea is to encode each continuous decision variable $x_i$ into 
the squared amplitude i.e. the measurement probability
of a single qubit state
\begin{equation}
    |\psi_i\rangle = \sqrt{1 - p_i}\,|0\rangle + \sqrt{p_i}\,|1\rangle,
    \quad p_i = |\langle 1|\psi_i\rangle|^2 \in [0, 1],
    \label{eq:encoding}
\end{equation}
from which $x_i$ is recovered via a decoding map $x_i = g_i(p_i)$, 
such as the bounded affine map $x_i = L_i + (U_i - L_i)p_i$. This 
replaces the $b_i$-qubit register of \eqref{eq:bitwidth} with a single 
parametrized qubit per variable, eliminating discretization error at 
the representation level. To recover encoded variable values, we propose 
using amplitude estimation \cite{brassard2002quantum} which estimates 
exactly this measurement probability $p_i$ rather than naive repeated 
sampling or quantum state tomography \cite{bermejo2022variational}, 
achieving near-optimal precision scaling for continuous-value readout. 
We analyze how amplitude-estimation error propagates through the decoding 
map to produce decision-variable error and ultimately objective-value 
error, yielding a distinct width-versus-precision tradeoff relative to 
discretized variational formulations.

\section{Background and Related Work}
\label{sec:background}

\subsection{Current Approaches}

As noted in Section~\ref{sec:intro} applying variational methods \cite{cerezo2021variational,
farhi2014qaoa} to continuous-variable objectives requires
discretizing each decision variable into a fixed-resolution
binary string, incurring qubit overhead that grows
logarithmically with required precision and introducing
irreducible discretization error. This is particularly
constraining on current NISQ hardware, where even
high-performing processors such as IBM's Heron QPUs offer
only 133--156 qubits \cite{ibmhardware2024}. Tan et al.\
\cite{tan2021qubit} and Glos et al.\ \cite{glos2022space}
propose qubit-efficient encodings for binary and combinatorial
optimization respectively, but neither addresses the
continuous-variable setting, and discretization error persists
in any fixed-resolution encoding.

\subsection{Quantum State Encodings}

The idea of encoding continuous information into the parameters of 
individual qubit states has appeared in related contexts. 
P\'{e}rez-Salinas et al.\ \cite{perez2020reuploading} show that a 
single qubit with repeated data re-uploading is sufficient to 
construct a universal quantum classifier, exploiting the continuous 
degrees of freedom of the Bloch sphere for data encoding in a 
variational circuit. While this work targets classification rather 
than optimization, it establishes that single-qubit continuous 
encoding is expressive within the standard qubit model.

Most directly related to our work, Bermejo and Or\'{u}s 
\cite{bermejo2022variational} propose encoding up to three 
continuous variables per qubit by exploiting all degrees of freedom 
of the Bloch sphere, two angles and one radius, combined with 
quantum state tomography for readout, in a variational framework 
for continuous optimization on standard qubits. In a companion 
paper \cite{bermejo2023nonorthogonal}, the same authors apply a 
related non-orthogonal encoding to discrete optimization, again 
using tomography for readout. Our framework differs from these 
approaches in two key respects. First, we restrict to encoding a 
single variable per qubit via the measurement probability $p_i = 
P(\text{outcome } 1)$, which yields a direct connection to amplitude 
estimation as a targeted readout primitive. Second, rather than 
quantum state tomography, which reconstructs the full single-qubit 
state and is more expensive than necessary when only the scalar 
measurement probability needs to be recovered, we propose amplitude 
estimation \cite{brassard2002quantum, grinko2021iqae} for readout. 
This achieves near-optimal precision scaling of 
$\mathcal{O}(1/\epsilon)$ up to logarithmic factors, compared to 
$\mathcal{O}(1/\epsilon^2)$ for naive repeated sampling, and 
directly improves the cost of refining encoded variable precision.

\section{Amplitude-Based Encoding and Framework}
\label{sec:framework}

\subsection{Single-Qubit Encoding}

We propose encoding each continuous decision variable $x_i$ into 
the measurement probability of a single qubit. Concretely, the 
$i$-th variable is represented by the pure qubit state as in 
\eqref{eq:encoding}, where $p_i = |\langle 1|\psi_i\rangle|^2 
\in [0,1]$ is the probability of obtaining outcome $1$ upon 
measuring in the computational basis. The continuous decision 
variable is then recovered via a general decoding map 
$x_i = g_i(p_i)$, where $g_i : [0,1] \to \mathcal{X}_i$ is 
chosen to match the domain $\mathcal{X}_i$ of variable $x_i$.

For bounded variables $x_i \in [L_i, U_i]$, the natural choice 
is the bounded affine decoding map
\begin{equation}
    g_i(p_i) = L_i + (U_i - L_i)\,p_i,
    \label{eq:affine_decode}
\end{equation}
which bijects $p_i \in [0,1]$ onto $x_i \in [L_i, U_i]$ without 
any discretization. Unlike the binary encoding of 
\eqref{eq:decoding}, this representation is continuous at the 
level of the qubit state --- any value $x_i \in [L_i, U_i]$ is 
exactly representable for some $p_i \in [0,1]$, and no 
discretization error is introduced at the encoding stage.

Other decoding maps are possible for different variable domains. 
For unbounded variables $x_i \in \mathbb{R}$, suitable choices 
include the logit map $g_i(p_i) = \log(p_i / (1-p_i))$, the 
scaled tangent map $g_i(p_i) = \tan(\pi(p_i - \tfrac{1}{2}))$, 
and the shifted arctangent hyperbolic map $g_i(p_i) = 
\operatorname{arctanh}(2p_i - 1)$, all of which map $(0,1)$ 
to $\mathbb{R}$. This flexibility is a further advantage over 
binary discretization, which always requires a finite range 
$[L_i, U_i]$ to be specified in advance, and wider ranges 
directly increase the qubit count through the logarithmic 
dependence of 
\eqref{eq:bitwidth}. Variables whose natural domain is unbounded 
or unknown must be artificially bounded before encoding, 
introducing an additional modeling assumption that also inflates 
qubit requirements. In the amplitude-based framework, unbounded 
domains are handled naturally by the choice of decoding map, 
without any prior range commitment or associated qubit overhead. The tradeoff is that all three maps diverge near the boundaries $p_i \to 0$ and 
$p_i \to 1$, introducing error amplification that must be accounted for in practice, which is beyond the scope of this paper. Thus, throughout we use the bounded affine map \eqref{eq:affine_decode} 
as the primary decoding map so that the precision with which $x_i$ can be recovered is determined 
entirely by how accurately $p_i$ can be estimated from 
measurements, which we address in Section~\ref{sec:readout}.

For an $n$-variable problem, the full solution $\mathbf{x} = 
(x_1, \ldots, x_n)$ is encoded across $n$ qubits, one per 
variable, yielding a total qubit cost of $\Theta(n)$ regardless 
of the desired precision. This contrasts with the discretized 
baseline, which requires $\sum_{i=1}^n b_i = \Theta(n \log(R/\eta))$ 
qubits for uniform range $R$ and precision $\eta$.

\subsection{Variational Circuit and Optimization Loop}

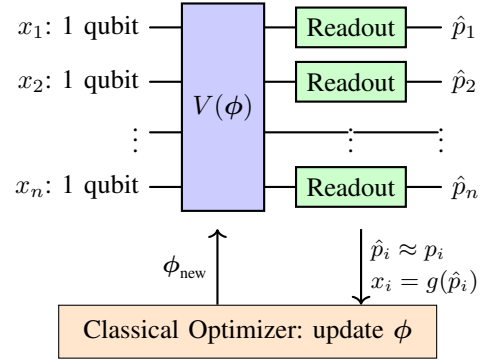
\begin{figure}[t]
\centering
\begin{tikzpicture}
    % Quantum circuit node
    \node[anchor=center] (circuit) {
        \begin{quantikz}[row sep=0.4em, column sep=1.2em]
            \lstick{$x_1$: 1 qubit} & \gate[4, style={fill=blue!20}]{V(\boldsymbol{\phi})} & \gate[style={fill=green!20}]{\text{Readout}} & \rstick{$\hat{p}_1$} \\
            \lstick{$x_2$: 1 qubit} & & \gate[style={fill=green!20}]{\text{Readout}} & \rstick{$\hat{p}_2$} \\
            \lstick{\vdots} & & \vdots & \vdots \\
            \lstick{$x_n$: 1 qubit} & & \gate[style={fill=green!20}]{\text{Readout}} & \rstick{$\hat{p}_n$}
        \end{quantikz}
    };
    % Classical optimizer box below
    \node[draw, rectangle, fill=orange!20, below=1.0cm of circuit, minimum width=5cm, minimum height=0.7cm] 
        (optimizer) {Classical Optimizer: update $\boldsymbol{\phi}$};
    % Right arrow: decoded values going down (right side)
    \draw[->, thick] 
        ([xshift=1.5cm]circuit.south) -- 
        node[right, align=left, font=\small] {$\hat{p}_i \approx p_i$\\[2pt]$x_i = g(\hat{p}_i)$} 
        ([xshift=1.5cm]optimizer.north);
    % Left arrow: new parameters going up (left side)
    \draw[->, thick] 
        ([xshift=-0.4cm]optimizer.north) -- 
        node[left, font=\small] {$\boldsymbol{\phi}_{\text{new}}$} 
        ([xshift=-0.4cm]circuit.south);
        
\end{tikzpicture}
\caption{Proposed amplitude-based variational circuit for $n$ 
continuous decision variables. Each variable $x_i$ is encoded 
into the measurement probability $p_i$ of a single qubit as in 
\eqref{eq:encoding}. The readout block estimates $p_i$; the 
choice of readout procedure and its precision scaling are 
discussed in Section~\ref{sec:readout}. Estimated probabilities 
$\hat{p}_i \approx p_i$ are decoded classically using a decoding 
map as $x_i = g(\hat{p}_i)$ and passed to a classical optimizer 
which updates $\boldsymbol{\phi}$ iteratively.}
\label{fig:proposed_circuit}
\end{figure}

The $n$ encoding qubits are initialized and jointly processed by 
a parametrized variational ansatz $V(\boldsymbol{\phi})$, where 
$\boldsymbol{\phi} \in \mathbb{R}^m$ are the trainable circuit 
parameters. After applying $V(\boldsymbol{\phi})$, the $i$-th 
qubit is in state $|\psi_i(\boldsymbol{\phi})\rangle$ with 
measurement probability
\begin{equation*}
    p_i(\boldsymbol{\phi}) = \langle \psi_i(\boldsymbol{\phi}) | 
    \Pi_1 | \psi_i(\boldsymbol{\phi}) \rangle, \quad 
    \Pi_1 = |1\rangle\langle 1|,
\end{equation*}
and the decoded variable is $x_i(\boldsymbol{\phi}) = L_i + 
(U_i - L_i)\,p_i(\boldsymbol{\phi})$. The objective function 
evaluated at the encoded point is
\begin{equation}
    F(\boldsymbol{\phi}) = f\!\left(x_1(\boldsymbol{\phi}), 
    \ldots, x_n(\boldsymbol{\phi})\right) = 
    f\!\left(\mathbf{g}(\mathbf{p}(\boldsymbol{\phi}))\right),
    \label{eq:objective}
\end{equation}
where $\mathbf{g} = (g_1, \ldots, g_n)$ is the vector of decoding 
maps. The variational optimization loop minimizes $F(\boldsymbol{\phi})$ 
over $\boldsymbol{\phi}$ via a classical outer optimizer:
\begin{equation*}
    \boldsymbol{\phi}^* = \arg\min_{\boldsymbol{\phi}} \; 
    F(\boldsymbol{\phi})
\end{equation*}
At each iteration, the quantum circuit prepares 
$V(\boldsymbol{\phi})|\mathbf{0}\rangle$, the measurement 
probabilities $\{p_i(\boldsymbol{\phi})\}$ are estimated via 
the readout procedure described in Section~\ref{sec:readout}, 
the decoded variables $\{x_i(\boldsymbol{\phi})\}$ are computed 
classically, the objective $F(\boldsymbol{\phi})$ is evaluated, 
and the classical optimizer updates $\boldsymbol{\phi}$. This 
hybrid loop continues until a convergence criterion is met. The 
resulting circuit structure is illustrated in 
Fig.~\ref{fig:proposed_circuit}, contrasting the single-qubit-per-variable 
encoding with the multi-qubit registers of Fig.~\ref{fig:discrete_circuit}.

% \begin{figure}[t]
% \centering
% \begin{tikzpicture}
%     % Quantum circuit node
%     \node[anchor=center] (circuit) {
%         \begin{quantikz}[row sep=0.4em, column sep=1.2em]
%             \lstick{$x_1$: 1 qubit} & \gate[4]{V(\boldsymbol{\phi})} & \gate{\text{Readout}} & \rstick{$\hat{p}_1$} \\
%             \lstick{$x_2$: 1 qubit} & & \gate{\text{Readout}} & \rstick{$\hat{p}_2$} \\
%             \lstick{\vdots} & & \vdots & \vdots \\
%             \lstick{$x_n$: 1 qubit} & & \gate{\text{Readout}} & \rstick{$\hat{p}_n$}
%         \end{quantikz}
%     };

%     % Classical optimizer box below
%     \node[draw, rectangle, below=1.0cm of circuit, minimum width=5cm, minimum height=0.7cm] 
%         (optimizer) {Classical Optimizer: update $\boldsymbol{\phi}$};

%     % Right arrow: decoded values going down (right side)
%     \draw[->, thick] 
%         ([xshift=1.5cm]circuit.south) -- 
%         node[right, align=left, font=\small] {$\hat{p}_i \approx p_i$\\[2pt]$x_i = g(\hat{p}_i)$} 
%         ([xshift=1.5cm]optimizer.north);

%     % Left arrow: new parameters going up (left side)
%     \draw[->, thick] 
%         ([xshift=-0.4cm]optimizer.north) -- 
%         node[left, font=\small] {$\boldsymbol{\phi}_{\text{new}}$} 
%         ([xshift=-0.4cm]circuit.south);
        
% \end{tikzpicture}

\section{Readout and Amplitude Estimation}
\label{sec:readout}

Once the variational circuit $V(\boldsymbol{\phi})$ has been applied,
each qubit $i$ is in a state $|\psi_i(\boldsymbol{\phi})\rangle$
whose measurement probability $p_i(\boldsymbol{\phi})$ encodes the
corresponding decision variable via the decoding map $g_i$. The
central readout task is to estimate $p_i$ accurately from quantum
measurements. We consider three approaches of different levels  of
sophistication and show that amplitude estimation should be the natural
readout primitive for this encoding.

\subsection{Naive Sampling}

The simplest strategy is to measure qubit $i$ repeatedly in the
computational basis and estimate $p_i$ by the empirical frequency
of outcome $1$. If $T$ independent shots are performed, the
estimator $\hat{p}_i = (\text{number of 1-outcomes})/T$ satisfies
the Chernoff--Hoeffding bound \cite{hoeffding1963}
\begin{equation*}
    \Pr\!\left[|\hat{p}_i - p_i| \geq \epsilon\right]
    \leq 2\exp(-2T\epsilon^2)
\end{equation*}
This gives us the shot requirement $T$ as
\begin{equation*}
    T = \mathcal{O}\!\left(\frac{1}{\epsilon^2}
    \log \frac{1}{\delta}\right)
\end{equation*}
to guarantee $|\hat{p}_i - p_i| \leq \epsilon$ with probability
at least $1 - \delta$. This $\mathcal{O}(1/\epsilon^2)$ scaling
is the standard shot-noise limit and represents the baseline
readout cost.

\subsection{Quantum State Tomography}

Single-qubit state tomography \cite{paris2004quantum} reconstructs
the full density matrix $\rho_i$ by measuring in multiple Pauli
bases ($X$, $Y$, $Z$) and extracts
$p_i = \langle 1|\rho_i|1\rangle$. While this yields the complete
state description, the additional information, relative phase
and purity, is unnecessary when only $p_i$ is needed for
decoding. Tomography still requires $\mathcal{O}(1/\epsilon^2)$
total shots for additive error $\epsilon$, offering no asymptotic
advantage over naive sampling while incurring basis-switching and
post-processing overhead. This is the readout strategy used by
Bermejo and Or\'{u}s \cite{bermejo2022variational}.

\subsection{Quantum Amplitude Estimation}

Quantum amplitude estimation (QAE) \cite{brassard2002quantum}
achieves a quadratic speedup by coherently amplifying the target
outcome probability, the quantity our encoding stores.

\subsubsection{Oracle and Grover Operator}

In our setting, the oracle $\mathcal{A}_i$= $V(\boldsymbol{\phi}))|_i$ for qubit $i$ is the action of $V(\boldsymbol{\phi})$ restricted to qubit $i$, preparing
$\mathcal{A}_i|0\rangle = |\psi_i(\boldsymbol{\phi})\rangle =
\sqrt{1-p_i}\,|0\rangle + \sqrt{p_i}\,|1\rangle$. The Grover
operator \cite{grover1996} is
$Q = -\mathcal{A}\,S_0\,\mathcal{A}^\dagger\,S_\chi$, where
$S_0 = I - 2|0\rangle\langle 0|$ and
$S_\chi = I - 2|1\rangle\langle 1|$. Acting on qubit $i$, these
reflections reduce to $S_0 = -Z$ and $S_\chi = Z$, so the minus
signs cancel and the Grover operator simplifies to
\begin{equation}
    Q_i = \mathcal{A}_i\, Z\, \mathcal{A}_i^\dagger\, Z
    \label{eq:grover_single}
\end{equation}
Geometrically, $Q_i$ rotates the state by $2\theta_i$ in the
real plane spanned by $|0\rangle$ and $|1\rangle$, where
$p_i = \sin^2(\theta_i)$. After $m$ applications,
\begin{equation*}
    Q_i^m \mathcal{A}_i |0\rangle =
    \cos\!\big((2m{+}1)\theta_i\big)|0\rangle +
    \sin\!\big((2m{+}1)\theta_i\big)|1\rangle,
\end{equation*}
so the probability of outcome $1$ oscillates as
$\sin^2\!\big((2m{+}1)\theta_i\big)$. Amplitude estimation
exploits this oscillatory dependence to infer $\theta_i$
and hence $p_i$.

\subsubsection{Standard QAE via Phase Estimation}

The original QAE algorithm \cite{brassard2002quantum} applies
quantum phase estimation \cite{cleve1998} to $Q_i$, extracting $\theta_i$ with
$\mathcal{O}(\epsilon^{-1} \log \delta^{-1})$ oracle calls, a quadratic improvement. However, it requires
$\mathcal{O}(\log 1/\epsilon)$ ancilla qubits and
controlled-$Q_i^{2^j}$ circuits, decreasing qubit efficiency, increasing depth, and making it impractical on near-term hardware.

\begin{algorithm}[t]
\caption{IQAE Readout for Single Qubit $i$}
\label{alg:iqae}
\begin{algorithmic}[1]
\Require Oracle $\mathcal{A}_i$=$V(\boldsymbol{\phi}))|_i$ (from $V(\boldsymbol{\phi})$ on
         qubit $i$)  , accuracy $\epsilon$, failure probability
         $\delta$, shots per round $N_{\mathrm{shots}}$
\Ensure $\hat{p}_i$ with
        $|\hat{p}_i - p_i| \leq \epsilon$,
        probability $\geq 1 - \delta$
\State $Q_i \gets \mathcal{A}_i\, Z\, \mathcal{A}_i^\dagger\, Z$
       \Comment{Eq.~\eqref{eq:grover_single}}
\State $[\theta_{\min}, \theta_{\max}] \gets [0,\, \pi/2]$
\While{$\theta_{\max} - \theta_{\min} > 2\arcsin\!\big(\sqrt{\epsilon}\,\big)$}
    \State $m_k \gets \left\lfloor
           \frac{\pi}{4(\theta_{\max} - \theta_{\min})}
           - \frac{1}{2} \right\rfloor$
    \For{$j = 1$ to $N_{\mathrm{shots}}$}
        \State Prepare $|0\rangle$; apply $\mathcal{A}_i$ then
               $Q_i^{m_k}$; measure $o_j \in \{0,1\}$
    \EndFor
    \State Compute Clopper--Pearson bounds on
           $\sin^2\!\big((2m_k{+}1)\theta_i\big)$
           from $\{o_j\}$ at confidence $1 - \delta_k$
    \State Map bounds to updated
           $[\theta_{\min}, \theta_{\max}]$
\EndWhile
\State \Return $\hat{p}_i \gets
       \sin^2\!\big((\theta_{\min} + \theta_{\max})/2\big)$
\end{algorithmic}
\end{algorithm}

\subsection{Iterative Quantum Amplitude Estimation}

Iterative QAE (IQAE) \cite{grinko2021iqae} replaces the QPE
circuit with an adaptive classical--quantum loop, maintaining a
confidence interval $[\theta_{\min}, \theta_{\max}]$ that is
progressively narrowed. At each round $k$, IQAE selects a Grover
depth $m_k$ such that $\sin^2\!\big((2m_k{+}1)\theta\big)$ remains
monotonic over the current interval, ensuring the measurement
outcome unambiguously tightens the bounds. After $N_{\mathrm{shots}}$
measurements at depth $m_k$, Clopper--Pearson confidence bounds
on
$\sin^2\!\big((2m_k{+}1)\theta_i\big)$ are mapped back to tighter
bounds on $\theta_i$. As the interval shrinks, $m_k$ grows. The procedure terminates
when the implied confidence interval for $p_i = \sin^2(\theta_i)$
has width at most $2\epsilon$. IQAE achieves
\begin{equation}
    T = \mathcal{O}\!\left(\frac{1}{\epsilon}
    \log \frac{1}{\delta}\right)
    \label{eq:iqae_complexity}
\end{equation}
total oracle calls, matching standard QAE, without any ancilla
qubits or controlled unitaries. Asymptotically optimal variants
using maximum likelihood estimation achieve the
information-theoretic lower bound
$T = \Theta\!\big(\epsilon^{-1} \log \delta^{-1}\big)$ with
provably minimal constant factors \cite{grinko2021iqae}; any of
these can serve as the primary readout primitive in our proposed framework.

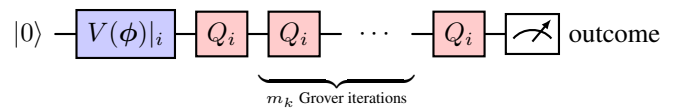
\begin{figure}[b]
\centering
\begin{quantikz}[row sep=0.5em, column sep=0.8em]
\lstick{$|0\rangle$} &
\gate[style={fill=blue!20}]{V(\boldsymbol{\phi})|_i} &
\gate[1, style={fill=red!20}]{Q_i} &
\gate[1, style={fill=red!20}]{Q_i} &
\midstick[1]{$\cdots$} &
\gate[1, style={fill=red!20}]{Q_i} &
\meter{}
\rstick[1]{outcome}
\end{quantikz}
\vspace{0.1em}
\centerline{\small $\underbrace{\hspace{6.5em}}_{m_k \text{ Grover iterations}}$}
\caption{IQAE circuit for qubit $i$ in round $k$. The oracle
$\mathcal{A}_i = V(\boldsymbol{\phi})|_i$ prepares 
$|\psi_i(\boldsymbol{\phi})\rangle$, followed by $m_k$ 
applications of $Q_i$ from \eqref{eq:grover_single}. 
Measurement outcomes update the confidence interval for 
$p_i = \sin^2(\theta_i)$.}
\label{fig:iqae_circuit}
\end{figure}

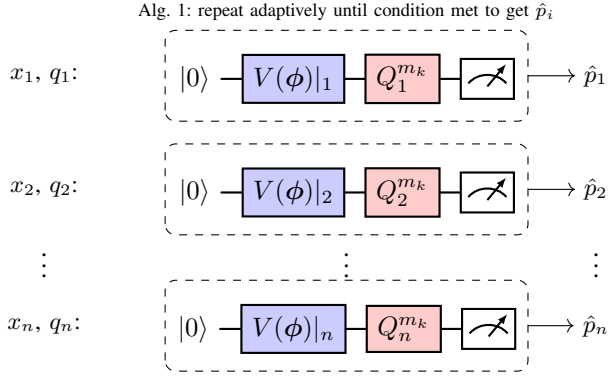
\begin{figure}[t]
\centering
\begin{tikzpicture}
    % Qubit 1
    \node[font=\small] (x1) at (-4.0, 2) {$x_1$, $q_1$:};
    \node (circ1) at (0, 2) {
        \begin{quantikz}[row sep=0.5em, column sep=0.8em]
            \lstick{$|0\rangle$} & \gate[style={fill=blue!20}]{V(\boldsymbol{\phi})|_1} & \gate[style={fill=red!20}]{Q_1^{m_k}} & \meter{}
        \end{quantikz}
    };
    \draw[dashed, rounded corners] (-2.4, 1.4) rectangle (2.4, 2.6);
    \node[font=\scriptsize] at (0, 2.85) {Alg.~1: repeat adaptively until condition met to get $\hat{p}_i$};
    \node[font=\small] (p1) at (3.3, 2) {$\hat{p}_1$};
    \draw[->] (2.4, 2) -- (p1);
    % Qubit 2
    \node[font=\small] (x2) at (-4.0, 0.5) {$x_2$, $q_2$:};
    \node (circ2) at (0, 0.5) {
        \begin{quantikz}[row sep=0.5em, column sep=0.8em]
            \lstick{$|0\rangle$} & \gate[style={fill=blue!20}]{V(\boldsymbol{\phi})|_2} & \gate[style={fill=red!20}]{Q_2^{m_k}} & \meter{}
        \end{quantikz}
    };
    \draw[dashed, rounded corners] (-2.4, -0.1) rectangle (2.4, 1.1);
    \node[font=\small] (p2) at (3.3, 0.5) {$\hat{p}_2$};
    \draw[->] (2.4, 0.5) -- (p2);
    % Vdots
    \node at (-4.0, -0.4) {$\vdots$};
    \node at (0, -0.4) {$\vdots$};
    \node at (3.3, -0.4) {$\vdots$};
    % Qubit n
    \node[font=\small] (xn) at (-4.0, -1.3) {$x_n$, $q_n$:};
    \node (circn) at (0, -1.3) {
        \begin{quantikz}[row sep=0.5em, column sep=0.8em]
            \lstick{$|0\rangle$} & \gate[style={fill=blue!20}]{V(\boldsymbol{\phi})|_n} & \gate[style={fill=red!20}]{Q_n^{m_k}} & \meter{}
        \end{quantikz}
    };
    \draw[dashed, rounded corners] (-2.4, -1.9) rectangle (2.4, -0.7);
    \node[font=\small] (pn) at (3.3, -1.3) {$\hat{p}_n$};
    \draw[->] (2.4, -1.3) -- (pn);
\end{tikzpicture}

\caption{Variational Ansatz with IQAE readout. For each variable $x_i$, 
qubit $q_i$ is prepared by the oracle 
$\mathcal{A}_i = V(\boldsymbol{\phi})|_i$ and the Grover 
operator $Q_i^{m_k}$ of \eqref{eq:grover_single} amplifies 
the target amplitude. The dashed box indicates that the entire 
prepare--amplify--measure cycle is repeated adaptively across 
rounds as specified in Algorithm~\ref{alg:iqae}, with $m_k$ 
increasing each round. Each qubit is read out independently; 
no ancilla qubits are required.}
\label{fig:iqae_readout}
\end{figure}

In our setting, $\mathcal{A}_i = V(\boldsymbol{\phi})|_i$
and the reflections in $Q_i$ of \eqref{eq:grover_single} reduce
to single-qubit $Z$ gates, so no ancilla qubits or controlled
unitaries beyond the ansatz itself are required.
The circuit for one IQAE round is shown in 
Fig.~\ref{fig:iqae_circuit} and Algorithm~\ref{alg:iqae} contains the complete readout procedure for a qubit. Fig.~\ref{fig:iqae_readout} shows the variational ansatz employing this algorithm.

\subsection{Readout Comparison}

Table~\ref{tab:readout} summarizes the readout strategies, where
query complexity counts the total number of applications of
$\mathcal{A}_i$ or $\mathcal{A}_i^\dagger$ across all circuit
executions. This is the standard complexity measure in the
amplitude estimation literature because it captures total
\emph{quantum work}: every oracle call requires running the state
preparation unitary $V(\boldsymbol{\phi})|_i$ on hardware, which
dominates the per-execution cost over ancillary operations like
the $Z$ gates in $Q_i$ or the classical post-processing. Counting
oracle calls therefore normalizes across methods that distribute
quantum work differently, naive sampling spreads
$\mathcal{O}(\epsilon^{-2})$ oracle calls across equally many
shallow circuit executions, while IQAE packs
$\mathcal{O}(\epsilon^{-1})$ oracle calls into far fewer but
progressively deeper circuits, with each round-$k$ execution
containing $2m_k + 1$ oracle calls. 
For the affine map \eqref{eq:affine_decode}, IQAE's
$\mathcal{O}(1/\epsilon)$ scaling translates directly to
$\mathcal{O}(1/\epsilon)$ cost for variable precision
$|\hat{x}_i - x_i| \leq (U_i - L_i)\epsilon$, as formalized in
Section~\ref{sec:error}.
\begin{table}[h]
\centering
\caption{Readout comparison for estimating $p_i$ to additive
accuracy $\epsilon$ with failure probability $\delta$.}
\label{tab:readout}
\begin{tabular}{lcc}
\hline
\textbf{Method} & \textbf{Query Complexity} & \textbf{Ancillae} \\
\hline
Naive sampling
    & $\mathcal{O}(\epsilon^{-2} \log \delta^{-1})$ & 0 \\
State tomography
    & $\mathcal{O}(\epsilon^{-2} \log \delta^{-1})$ & 0 \\
Standard QAE
    & $\mathcal{O}(\epsilon^{-1} \log \delta^{-1})$
    & $\mathcal{O}(\log \epsilon^{-1})$ \\
IQAE 
    & $\mathcal{O}(\epsilon^{-1} \log \delta^{-1})$ & 0 \\
\hline
\end{tabular}
\end{table}

\section{Error Propagation and Resource Tradeoff}
\label{sec:error}

We now analyze how readout error propagates to decision-variable
error and then to objective-value error in the discretized and amplitude-based frameworks,
establishing a resource comparison on common footing.

\subsection{Variable-Level Precision}

\subsubsection{Discretized Baseline}

As established in Section~\ref{sec:intro}, each variable
$x_i \in [L_i, U_i]$ encoded into $b_i$ qubits via
\eqref{eq:decoding} incurs an irreducible discretization error
$|\tilde{x}_i - x_i^*| \leq R_i / 2^{b_i}$ that can only be
reduced by adding qubits. Each additional bit of precision
costs one qubit per variable.

\subsubsection{Amplitude-Based Framework}

Each variable is encoded into the measurement probability $p_i$
of a single qubit. IQAE estimates $p_i$ to additive accuracy
$\epsilon_i$, yielding variable error
\begin{equation*}
    |\hat{x}_i - x_i| = R_i\,|\hat{p}_i - p_i|
    \leq R_i\,\epsilon_i
\end{equation*}
through the affine map \eqref{eq:affine_decode}. To achieve
variable precision $\eta_i$, we set $\epsilon_i = \eta_i / R_i$,
requiring $\mathcal{O}(R_i/\eta_i \cdot \log 1/\delta)$ oracle
calls via \eqref{eq:iqae_complexity} on a single qubit. There
is no discretization floor: precision is limited only by the
estimation budget and can be improved without adding qubits.

\subsubsection{Comparison}

For uniform range $R$ and target variable precision $\eta$, the
discretized baseline requires $\lceil \log_2(R/\eta) \rceil$
qubits per variable, totaling
$\Theta(n \log(R/\eta))$ qubits with an irreducible error
floor at $\eta$. Our framework requires one qubit per variable,
totaling $\Theta(n)$ qubits, and achieves the same precision
via $\mathcal{O}(R/\eta \cdot \log 1/\delta)$ oracle calls
per variable. The tradeoff is clear: the discretized approach
pays for precision in qubit width; our approach pays in
estimation depth.

\subsection{Objective-Level Precision}

We propagate variable-level errors through the objective
function $f$ to compare total per-iteration costs for achieving
a target objective error $\delta_F$.

\subsubsection{Lipschitz Assumption}

We assume $f$ is $K$-Lipschitz continuous with respect to the
$\ell_\infty$ norm:
\begin{equation*}
    |f(\mathbf{y}) - f(\mathbf{x})| \leq K \|\mathbf{y} -
    \mathbf{x}\|_\infty \quad \forall\; \mathbf{x}, \mathbf{y}
    \in \textstyle\prod_{i=1}^n [L_i, U_i].
\end{equation*}
The $\ell_\infty$ Lipschitz condition bounds the objective error
by the worst-case coordinate error rather than the aggregate
$\ell_2$ norm, yielding per-variable precision requirements
that do not degrade with problem dimension $n$.
This condition is equivalent to requiring bounded partial
derivatives $|\partial f / \partial x_i| \leq K$ for all $i$,
and is satisfied by the broad class of objectives with bounded
gradients, including separable, additively structured, and
mildly coupled functions common in engineering and finance
applications.

\subsubsection{Discretized Baseline}

The objective error has two independent sources. First,
discretization error propagates through $f$:
\begin{equation*}
    |f(\tilde{\mathbf{x}}) - f(\mathbf{x}^*)| \leq
    K \max_i \frac{R_i}{2^{b_i}} \leq \frac{KR}{2^b}
\end{equation*}
for uniform $R$ and $b$. To make this at most $\delta_F$
requires $b = \mathcal{O}(\log(KR/\delta_F))$ qubits
per variable. Second, the classical optimizer estimates the
objective by averaging $f(\tilde{\mathbf{x}}^{(t)})$ over $T$
sampled bitstrings per iteration. Each shot yields a different
bitstring and hence a different decoded point, introducing
shot-noise variance scaling as $\mathcal{O}(1/T)$. Achieving
objective precision $\delta_F$ from shot noise alone requires
$T = \mathcal{O}(1/\delta_F^2)$ circuit executions per
iteration. Each circuit execution yields all $n$
decoded variables simultaneously.

\subsubsection{Amplitude-Based Framework}

In our framework, the objective error arises solely from IQAE
estimation errors propagated through $f$:
\begin{equation*}
    |f(\hat{\mathbf{x}}) - f(\mathbf{x})| \leq
    K \|\hat{\mathbf{x}} - \mathbf{x}\|_\infty
    = K \max_i R_i \epsilon_i \leq KR\,\epsilon
\end{equation*}
for uniform $R$ and $\epsilon$. Setting this to $\delta_F$
gives $\epsilon = \delta_F/(KR)$, independent of $n$. Each
variable's IQAE then requires
$\mathcal{O}(KR/\delta_F)$ oracle calls.

Since each qubit's IQAE runs at a different adaptive Grover
depth $m_k$, and the Grover operator $Q_i$ of
\eqref{eq:grover_single} involves the full ansatz
$V(\boldsymbol{\phi})$ and its inverse when the ansatz
contains entangling gates, each IQAE circuit execution
requires running the complete $n$-qubit ansatz. The
differing Grover depths prevent batching multiple qubits'
IQAE rounds into a single circuit execution. The total
across $n$ variables is therefore
\begin{equation*}
    \mathcal{O}\!\left(\frac{nKR}{\delta_F}
    \log \frac{1}{\delta}\right)
\end{equation*}
circuit executions / oracle calls per iteration. The factor of $n$ arises
because each of the $n$ variables requires its own sequence
of IQAE rounds, each involving the full ansatz as the oracle.
There is no discretization error and no additional shot-noise
term.

\subsection{Resource Comparison}

Table~\ref{tab:resource} compares the two frameworks for
achieving objective precision $\delta_F$.

\begin{table}[h]
\centering
\caption{Per-iteration resource comparison for target objective
precision $\delta_F$ with $n$ variables, uniform range $R$,
and $K$-Lipschitz ($\ell_\infty$) objective. Log factors in
$\delta$ suppressed.}
\label{tab:resource}
\begin{tabular}{lcc}
\hline
 & \textbf{Discretized} & \textbf{Ours (IQAE)} \\
\hline
Qubits
    & $\mathcal{O}\!\left(n \log \frac{KR}{\delta_F}\right)$
    & $n$ \\[4pt]
Circuit exec./ Oracle calls
    & $\mathcal{O}\!\left(\delta_F^{-2}\right)$
    & $\mathcal{O}\!\left(\frac{nKR}{\delta_F}\right)$ \\[4pt]
Error sources
    & discretization $+$ shot noise
    & estimation only \\[2pt]
Precision floor
    & yes ($R/2^{b}$)
    & none \\
\hline
\end{tabular}
\end{table}

The two frameworks exhibit complementary scaling. In qubit
count, our framework requires $n$ qubits regardless of
precision, versus $\mathcal{O}(n \log(KR/\delta_F))$
for the discretized baseline, a saving that grows as
finer precision is demanded. In circuit executions per
iteration, our framework scales as
$\mathcal{O}(1/\delta_F)$ versus $\mathcal{O}(1/\delta_F^2)$,
a quadratic improvement inherited from amplitude estimation.
The cost of this improvement is the $nKR$ prefactor in
our circuit execution count: the factor of $n$ arises from
sequential per-variable IQAE readout using the full
entangling ansatz as the oracle, while $KR$ reflects the
Lipschitz-mediated precision requirement. Our framework
requires fewer circuit executions when
$\delta_F < 1/(nKR)$.

\section{Numerical Illustration}

To make the comparison concrete, Table~\ref{tab:numerical}
instantiates the resource costs for a representative
problem with $n = 10$ variables, range $R = 5$ (e.g.,
each $x_i \in [-2.5, 2.5]$), and Lipschitz constant $K = 2$,
corresponding to a mildly coupled objective with bounded
partial derivatives. The crossover point is
$\delta_F = 1/(nKR) = 0.01$.

\begin{table}[h]
\centering
\caption{Concrete resource costs for $n = 10$, $R = 5$,
$K = 2$. Discretized qubit count is
$10 \lceil \log_2(10/\delta_F) \rceil$; circuit executions
are $\lceil 1/\delta_F^2 \rceil$ (discretized) and
$\lceil nKR/\delta_F \rceil$ (ours).}
\label{tab:numerical}
\begin{tabular}{rcccc}
\hline
$\delta_F$ & \multicolumn{2}{c}{\textbf{Discretized}}
           & \multicolumn{2}{c}{\textbf{IQAE}} \\
 & Qubits & Circ.\ exec.
 & Qubits & Circ.\ exec.\\
\hline
$10^{-1}$ & 70  & $10^{2}$   & 10 & $10^{3}$ \\
$10^{-2}$ & 100 & $10^{4}$   & 10 & $10^{4}$ \\
$10^{-3}$ & 140 & $10^{6}$   & 10 & $10^{5}$ \\
$10^{-4}$ & 170 & $10^{8}$   & 10 & $10^{6}$ \\
\hline
\end{tabular}
\end{table}

At $\delta_F = 10^{-1}$, the discretized baseline uses
fewer circuit executions but already requires $70$ qubits
compared to $10$. At the crossover $\delta_F = 10^{-2}$,
circuit executions are comparable while the qubit gap widens
to $100$ versus $10$. Beyond this point, each additional
digit of precision costs $100\times$ more circuit executions
for the discretized baseline but only $10\times$ more for
our framework, while the qubit count continues to grow
for the baseline and remains fixed for ours. At
$\delta_F = 10^{-3}$, the discretized baseline requires
$140$ qubits, approaching the capacity of current
hardware such as IBM's $133$--$156$ qubit Heron
processors, whereas IQAE
remains at $10$ qubits with $10\times$ fewer circuit
executions.

On current NISQ hardware, qubit count is the binding
constraint. Our framework shifts the precision cost to
circuit executions, a resource that scales with wall-clock
time rather than hardware size. Additionally, it
eliminates the irreducible discretization floor: precision can
always be improved by investing additional circuit executions
without modifying the circuit structure or adding qubits.

\section{Conclusion}
\label{sec:conclusion}

We have proposed a qubit-efficient variational framework for
continuous optimization on standard gate-based quantum computers
that encodes each decision variable into the measurement
probability of a single qubit, replacing the multi-qubit binary
registers required by discretized approaches. By using iterative
quantum amplitude estimation for readout, the framework achieves
$\mathcal{O}(1/\delta_F)$ precision scaling in circuit executions
versus $\mathcal{O}(1/\delta_F^2)$ for the discretized baseline,
while reducing the qubit count from
$\mathcal{O}(n \log(KR/\delta_F))$ to $n$ and eliminating the
irreducible discretization error floor entirely. For a
representative $10$-variable problem, the discretized baseline
approaches the qubit capacity of current hardware at moderate
precision targets where our framework requires only $10$ qubits.

Continuous-variable quantum optimization has previously required
dedicated photonic or bosonic hardware to avoid discretization.
Our framework achieves this same discretization-free property
on standard qubit architectures, making it compatible with
the superconducting, trapped-ion, and neutral-atom platforms
where the majority of current quantum computing development
is concentrated. At the same time, by operating within the
qubit circuit model, the framework can leverage the entangling
ansatz structures, error mitigation techniques, and classical
optimizer integrations that have been developed for variational
quantum algorithms over the past decade. This positions
amplitude-based encoding as a bridge between the
discretization-free expressiveness of continuous-variable
methods and the hardware maturity of qubit-based platforms,
offering a new point in the width-versus-depth tradeoff space
for variational continuous optimization on near-term quantum
hardware.
\clearpage
\bibliographystyle{IEEEtran}
\bibliography{references}

\end{document}